\documentclass[conference]{IEEEtran}
\IEEEoverridecommandlockouts

\usepackage{cite}
\usepackage{amsmath,amssymb,amsfonts}
\usepackage{algorithmic}
\usepackage{graphicx}
\usepackage{textcomp}
\usepackage{xcolor}
\usepackage{tabularx}
\usepackage{url}
\usepackage{multirow}
\usepackage{enumitem}
\usepackage{caption}
\usepackage{cleveref}
\usepackage{booktabs}
\def\BibTeX{{\rm B\kern-.05em{\sc i\kern-.025em b}\kern-.08em
    T\kern-.1667em\lower.7ex\hbox{E}\kern-.125emX}}
\usepackage{pgfplots}
\pgfplotsset{compat=1.18}
\usepackage{pgfplotstable}
\usepackage{subcaption}
\crefname{figure}{Fig.}{Figs.} 
\Crefname{figure}{Fig.}{Figs.}
\begin{document}

\title{Revealing the Role of Audio Channels in ASR Performance Degradation\\
{\footnotesize }
\thanks{}
}

\author{\IEEEauthorblockN{Kuan-Tang Huang\IEEEauthorrefmark{1},
Li-Wei Chen\IEEEauthorrefmark{2}\IEEEauthorrefmark{4},
Hung-Shin Lee\IEEEauthorrefmark{4}, 
Berlin Chen\IEEEauthorrefmark{1}, and
Hsin-Min Wang\IEEEauthorrefmark{3}}
\IEEEauthorblockA{\IEEEauthorrefmark{1}Dept. Computer Science and Information Engineering, National Taiwan Normal University, Taiwan }
\IEEEauthorblockA{\IEEEauthorrefmark{2}Dept. Computer Science and Information Engineering, National Tsing Hua University, Taiwan
}
\IEEEauthorblockA{\IEEEauthorrefmark{3}Institute of Computer Science, Academia Sinica, Taiwan
}
\IEEEauthorblockA{\IEEEauthorrefmark{4}United Link Co., Ltd., Taiwan
}
}

\maketitle

\begin{abstract}
Pre-trained automatic speech recognition (ASR) models have demonstrated strong performance on a variety of tasks.
However, their performance can degrade substantially when the input audio comes from different recording channels.
While previous studies have demonstrated this phenomenon, it is often attributed to the mismatch between training and testing corpora.
This study argues that variations in speech characteristics caused by different recording channels can fundamentally harm ASR performance.
To address this limitation, we propose a normalization technique designed to mitigate the impact of channel variation by aligning internal feature representations in the ASR model with those derived from a clean reference channel.
This approach significantly improves ASR performance on previously unseen channels and languages, highlighting its ability to generalize across channel and language differences.
\end{abstract}


\begin{figure*}[t]
\centering
\includegraphics[width=0.85\linewidth]{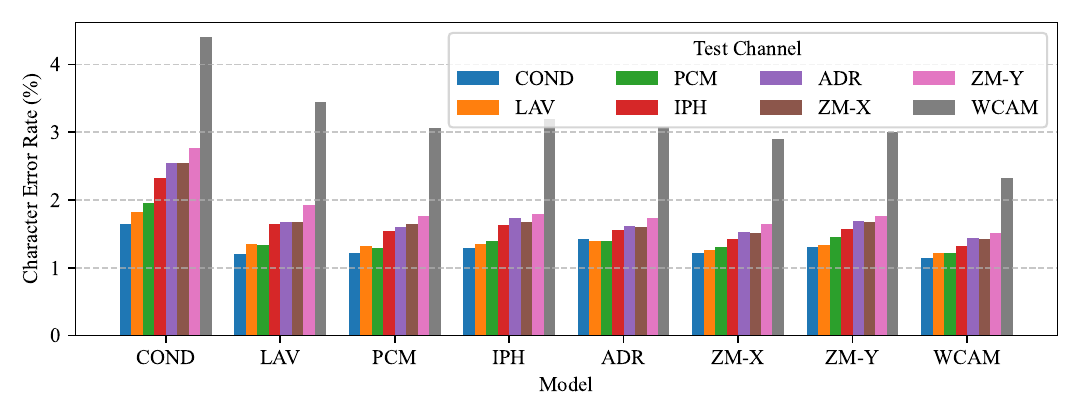}
\vspace{-10pt}
\caption{CERs (\%) across Different Channels and Test Channels. 
Each cluster of bars on the x-axis represents a specific fine-tuned model, and within each cluster, individual bars denote the CER achieved on different test channels.
The channels are abbreviated as follows: COND for Condenser, LAV for Lavalier, PCM for PC-Mic, IPH for iPhone, ADR for Android phone, ZM-X for ZOOM-X, ZM-Y for ZOOM-Y, and WCAM for Webcam.
}
\label{fig:channel-analysis}
\vspace{-15pt}
\end{figure*}

\section{Introduction}
Recent advances in automatic speech recognition (ASR) \cite{gulati2020,abouelenin2025,zhang2022,baevski2020} have been propelled by the development of large-scale pre-trained models.
These models, trained on extensive and diverse datasets, have enabled significant performance improvements in a wide range of downstream tasks and conditions.
A notable example is Whisper \cite{radford2023}, an open-source model trained on more than 680,000 hours of multilingual and multitask data, which exhibits considerable robustness across various domains and languages.
Similarly, SpeechStew \cite{chan2021} employs a mixture-of-corpora training strategy, utilizing diverse English datasets to cultivate general-purpose ASR capabilities.
Another contemporary model, the Universal Speech Model (USM) \cite{zhang2023} developed by Google, extends ASR training to more than 100 languages and domains via a unified encoder-decoder architecture.
These pre-trained ASR models are generally considered robust and effective under varied conditions, covering different speakers, domains, and noisy environments \cite{shraddha2022,wang_dysarthric_2024,qin2021,zusag2024}.
However, their performance may exhibit notable variation when evaluated using audio from different recording channels, stemming from variations in microphones or device configurations.

This performance variability presents a significant challenge for real-world applications, where ASR systems are frequently deployed in diverse acoustic environments and utilize a wide array of hardware.
For instance, speech recognition accuracy can degrade substantially when transitioning from high-quality studio microphones to consumer-grade mobile devices.
Such inconsistencies diminish the reliability and user experience of ASR-powered applications, including virtual assistants, transcription services, and accessibility tools, thus establishing channel robustness as a critical research objective.

Previous research \cite{wang2025} has predominantly framed this issue as one of domain mismatch \cite{mani2020,zhou2021,liao2019}---a discrepancy between the data distributions of training and testing channels---and proposed solutions involving data augmentation to simulate target channel characteristics during training.
This study extends beyond the conventional domain mismatch paradigm, presenting empirical evidence that intrinsic signal differences imparted by the recording channels are the main contributors to ASR performance degradation.
Our controlled experiments demonstrate a consistent performance hierarchy among channels, irrespective of the specific channel data used for fine-tuning.
This observation suggests that ASR performance degradation is influenced more significantly by fixed, channel-specific signal properties than by domain mismatch.
This finding is further elaborated in \Cref{sec:channel-observation}.

An intuitive way to address channel-induced signal distortions is to apply speech enhancement (SE) technology \cite{schroter2022,defossez2020,pandey2019}.
However, SE methods are widely documented to introduce processing artifacts that can adversely affect ASR performance \cite{ochiai2024,iwamoto2022,wang2024,ho2023,iwamoto2024}, making them less appropriate to improve channel robustness in this context.
Consequently, SE-based strategies are not investigated herein.
Instead, we introduce a novel methodology that aligns internal ASR feature representations with those derived from a clean reference channel.
This is accomplished by integrating lightweight adapter layers \cite{houlsby2019,hu2023,sung2022} into the encoder of a pre-trained ASR model and exclusively training these adapters to normalize intermediate features towards a clean-channel distribution.
This modular architecture facilitates the interchange of encoder modules at inference time without necessitating modifications to the decoder or other model components.
This approach yields notable performance improvements across diverse channels, including those not encountered during training, and demonstrates robust generalization capabilities.
Furthermore, optional fine-tuning of the decoder can further enhance performance, providing enhanced adaptability to various acoustic conditions.
Although trained on a single language, our modular encoder demonstrates consistent performance gains when evaluated on a different language, suggesting the potential for cross-lingual robustness.

Our contributions are summarized as follows:
\begin{enumerate}[noitemsep,leftmargin=*]
\item \textbf{Re-evaluating the impact of recording channels on ASR performance.}
We extend beyond the prevalent domain mismatch explanation, presenting empirical evidence that intrinsic factors induced by recording channels---such as microphone characteristics and acoustic distortions---are significant contributors to ASR performance degradation, frequently outweighing domain-specific effects.
\item \textbf{A modular normalization technique for enhanced channel robustness.}
We introduce an innovative normalization technique that transforms internal ASR representations to approximate those of clean-channel features, thereby enabling robust performance across diverse acoustic conditions.
The independent training of a modular encoder facilitates flexible integration with various decoders during inference and yields substantial performance improvements.
Additional performance enhancements can be realized through optional decoder fine-tuning, although the proposed method demonstrates efficacy even in its absence.
\end{enumerate}

\section{Proposed Method}

This section first presents an empirical analysis of the impact of recording channels on ASR performance using Whisper.
Subsequently, based on these observations, a novel normalization technique is proposed to mitigate channel-related performance degradation.

\begin{table*}[t]
\caption{CER (\%) and  Relative Improvement Rate (\%) of $\mathit{{Van}_{pre}}$ vs. $\mathit{{Van}_{adp}}$ on HAT.}
\label{tab:hat-result}
\centering
\small
\selectfont
\setlength{\tabcolsep}{3.75pt}
\begin{tabular}{l l *{8}{r r} r r}
\toprule
\multirow{2}{*}{Method} 
  & \multirow{2}{*}{Channel}
  & \multicolumn{2}{c}{COND}
  & \multicolumn{2}{c}{ADR}
  & \multicolumn{2}{c}{ZM-X}
  & \multicolumn{2}{c}{ZM-Y}
  & \multicolumn{2}{c}{IPH}
  & \multicolumn{2}{c}{LAV}
  & \multicolumn{2}{c}{PCM}
  & \multicolumn{2}{c}{WCAM}
  & \multicolumn{2}{c}{AVG} \\
\cmidrule(lr){3-4}
\cmidrule(lr){5-6}
\cmidrule(lr){7-8}
\cmidrule(lr){9-10}
\cmidrule(lr){11-12}
\cmidrule(lr){13-14}
\cmidrule(lr){15-16}
\cmidrule(lr){17-18}
\cmidrule(lr){19-20}
& & CER & rel. & CER & rel. & CER & rel. & CER & rel. & CER & rel. & CER & rel. & CER & rel. & CER & rel. & CER & rel. \\
\midrule
\multicolumn{20}{c}{\textbf{Upper half: Decoders trained on single channel data}} \\
\midrule
$\mathit{{Van}_{pre}}$ & COND & \textbf{1.64} & -- & 2.54 & -- & 2.55 & -- & 2.77 & -- & 2.32 & -- & 1.82 & -- & 1.96 & -- & 4.40 & -- & 2.50 & -- \\
$\mathit{{Van}_{adp}}$ & COND & 1.67 & -1.8 & \textbf{1.99} & 21.7 & \textbf{2.07} & 18.8 & \textbf{2.28} & 17.7 & \textbf{1.93} & 16.8 & \textbf{1.77} & 2.7 & \textbf{1.76} & 10.2 & \textbf{4.02} & 8.6 & \textbf{2.19} & 12.4 \\
\midrule
$\mathit{{Van}_{pre}}$ & ADR & 1.42 & -- & 1.62 & -- & 1.60 & -- & 1.74 & -- & 1.56 & -- & 1.39 & -- & 1.40 & -- & 3.08 & -- & 1.73 & -- \\
$\mathit{{Van}_{adp}}$ & ADR & \textbf{1.31} & 7.7 & \textbf{1.46} & 9.9 & \textbf{1.41} & 11.9 & \textbf{1.49} & 14.4 & \textbf{1.43} & 8.3 & \textbf{1.27} & 8.6 & \textbf{1.26} & 10.0 & \textbf{2.76} & 10.4 & \textbf{1.55} & 10.4 \\
\midrule
$\mathit{{Van}_{pre}}$ & ZM-X & 1.22 & -- & \textbf{1.53} & -- & 1.51 & -- & 1.65 & -- & 1.43 & -- & 1.27 & -- & 1.30 & -- & 2.90 & -- & 1.60 & -- \\
$\mathit{{Van}_{adp}}$ & ZM-X & \textbf{1.20} & 1.6 & 1.72 & -12.4 & \textbf{1.36} & 9.9 & \textbf{1.44} & 12.7 & \textbf{1.32} & 7.7 & \textbf{1.25} & 1.6 & \textbf{1.25} & 3.8 & \textbf{2.45} & 15.5 & \textbf{1.50} & 6.3 \\
\midrule
$\mathit{{Van}_{pre}}$ & ZM-Y & 1.31 & -- & 1.69 & -- & 1.67 & -- & 1.77 & -- & 1.58 & -- & \textbf{1.33} & -- & 1.45 & -- & 3.00 & -- & 1.73 & -- \\
$\mathit{{Van}_{adp}}$ & ZM-Y & \textbf{1.28} & 2.3 & \textbf{1.51} & 10.7 & \textbf{1.56} & 6.6 & \textbf{1.67} & 5.6 & \textbf{1.56} & 1.3 & 1.48 & -11.3 & \textbf{1.31} & 9.7 & \textbf{2.77} & 7.7 & \textbf{1.64} & 5.2 \\
\midrule
$\mathit{{Van}_{pre}}$ & IPH & 1.29 & -- & 1.73 & -- & 1.68 & -- & 1.79 & -- & 1.63 & -- & 1.35 & -- & 1.39 & -- & 3.19 & -- & 1.76 & -- \\
$\mathit{{Van}_{adp}}$ & IPH & \textbf{1.27} & 1.6 & \textbf{1.54} & 11.0 & \textbf{1.41} & 16.1 & \textbf{1.53} & 14.5 & \textbf{1.42} & 12.9 & \textbf{1.31} & 3.0 & \textbf{1.32} & 5.0 & \textbf{2.79} & 12.5 & \textbf{1.57} & 10.8 \\
\midrule
$\mathit{{Van}_{pre}}$ & LAV & \textbf{1.21} & -- & 1.68 & -- & 1.68 & -- & 1.92 & -- & 1.64 & -- & 1.35 & -- & 1.34 & -- & 3.44 & -- & 1.78 & -- \\
$\mathit{{Van}_{adp}}$ & LAV & 1.22 & -0.8 & \textbf{1.49} & 11.3 & \textbf{1.48} & 11.9 & \textbf{1.64} & 14.6 & \textbf{1.44} & 12.2 & \textbf{1.27} & 5.9 & \textbf{1.27} & 5.2 & \textbf{2.95} & 14.2 & \textbf{1.60} & 10.1 \\
\midrule
$\mathit{{Van}_{pre}}$ & PCM & 1.22 & -- & 1.60 & -- & 1.64 & -- & 1.77 & -- & 1.54 & -- & 1.32 & -- & 1.29 & -- & 3.07 & -- & 1.68 & -- \\
$\mathit{{Van}_{adp}}$ & PCM & \textbf{1.21} & 0.8 & \textbf{1.41} & 11.9 & \textbf{1.40} & 14.6 & \textbf{1.47} & 16.9 & \textbf{1.39} & 9.7 & \textbf{1.21} & 8.3 & \textbf{1.26} & 2.3 & \textbf{2.76} & 10.1 & \textbf{1.51} & 10.1 \\
\midrule
$\mathit{{Van}_{pre}}$ & WCAM & \textbf{1.14} & -- & 1.44 & -- & 1.42 & -- & 1.51 & -- & 1.32 & -- & 1.22 & -- & 1.22 & -- & 2.32 & -- & 1.45 & -- \\
$\mathit{{Van}_{adp}}$ & WCAM & \textbf{1.14} & 0.0 & \textbf{1.28} & 11.1 & \textbf{1.22} & 14.1 & \textbf{1.32} & 12.6 & \textbf{1.26} & 4.5 & \textbf{1.15} & 5.7 & \textbf{1.15} & 5.7 & \textbf{1.98} & 14.7 & \textbf{1.31} & 9.7 \\
\midrule
\multicolumn{20}{c}{\textbf{Lower half: Decoders trained exclude WCAM channel}} \\
\midrule
$\mathit{{Van}_{pre}}$ & \raisebox{-0.7ex}{\textasciitilde}WCAM & 1.04 & -- & 1.29 & -- & 1.27 & -- & 1.38 & -- & 1.25 & -- & 1.12 & -- & 1.08 & -- & 2.48 & -- & 1.36 & -- \\
$\mathit{{Van}_{adp}}$ & \raisebox{-0.7ex}{\textasciitilde}WCAM & \textbf{1.03} & 1.0 & \textbf{1.14} & 11.6 & \textbf{1.13} & 11.0 & \textbf{1.20} & 13.0 & \textbf{1.13} & 9.6 & \textbf{1.03} & 8.0 & \textbf{1.05} & 2.8 & \textbf{2.17} & 12.5 & \textbf{1.24} & 8.8 \\
\bottomrule
\end{tabular}
\vspace{-15pt}
\end{table*}

\begin{figure}[t]
\centering
\includegraphics[width=1\linewidth]{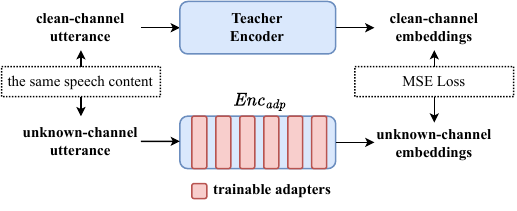}
\vspace{-15pt}
\caption{Training pipeline for the adapter-enhanced encoder.
}
\label{fig:adapter-training}
\vspace{-15pt}
\end{figure}

\subsection{Empirical Analysis of Channel Impact}
\label{sec:channel-observation}

To precisely evaluate the influence of recording channels, the $\mathit{{Whisper}_{small}}$ model was fine-tuned on data from each individual channel, and the resultant models were subsequently evaluated across all available channels.
The Hakka Across Taiwan (HAT) corpus \cite{liao2023}, which comprises simultaneous multi-channel recordings from eight distinct channels, was utilized for this purpose.
This experimental design effectively controls for variations attributable to speakers and linguistic content, thereby isolating the impact of channel differences on ASR performance.
Fine-tuning is necessitated by the underrepresentation of Hakka in Whisper's original training dataset.

As illustrated in \Cref{fig:channel-analysis}, performance trends exhibit consistency irrespective of the channel used for fine-tuning: channels that yield superior performance do so across all evaluated models, whereas channels yielding inferior performance consistently underperform.
This observation indicates that ASR performance is primarily governed by intrinsic signal characteristics introduced by each recording channel---such as microphone specifications, placement geometry, and acoustic distortions---rather than by mismatches between the domains of the training and testing data.
Were domain mismatch the predominant factor, it would be expected that each model would perform optimally on the specific channel on which it was fine-tuned; however, this outcome is not observed.

\subsection{Channel Normalization Technique}

Motivated by the observation of significant channel-specific effects on ASR performance, we propose a channel normalization technique designed to transform feature representations from disparate recording channels into a canonical, clean-channel feature space.
Our methodology leverages the established capabilities of pre-trained models by inserting adapter layers into the encoder and fine-tuning only these adapter modules, as depicted in \Cref{fig:adapter-training}.
The original pre-trained encoder serves as a teacher model, and our adapter-enhanced encoder, denoted as $\mathit{{Enc}_{adp}}$, is initialized with identical weights.
During training, utterances of the same speech content captured concurrently by multiple devices are used: a clean-channel utterance is input to the teacher encoder, while the corresponding utterance from various other channels is processed by $\mathit{{Enc}_{adp}}$.
The $\mathit{{Enc}_{adp}}$ module is trained by minimizing the mean squared error (MSE) between its output embeddings and those of the teacher model at the final encoder layer.
This training objective encourages $\mathit{{Enc}_{adp}}$ to normalize features towards the clean-channel feature space.
Although the MSE loss is computed solely at the final encoder layer, the adapter modules, integrated at multiple intermediate layers, facilitate progressive adjustment and normalization of features throughout the encoding process.
This multi-layer architecture promotes the progressive refinement of channel-invariant representations across different levels of abstraction, thereby supporting effective learning without necessitating explicit supervision at each intermediate layer.

A key property of our training data is that it contains the same speech content across different channels.
This ensures that the model learns to normalize variations specifically caused by channel differences, without conflating them with linguistic or other unrelated variations.
Additionally, inputting clean-channel utterance to 
$\mathit{{Enc}_{adp}}$ enables the model to preserve its original embeddings when the input is already of high quality, thereby maintaining performance when normalization is superfluous. In our experiments, the input to $\mathit{{Enc}_{adp}}$ encompasses seven distinct recording conditions, including the clean channel.
This comprehensive channel diversity exposes the model to a wide spectrum of acoustic characteristics, thereby enhancing its generalization capabilities and mitigating the risk of overfitting to specific channels.

Furthermore, as the adapter is trained without explicit channel labels, it learns to detect and compensate for channel-specific distortions directly from the acoustic input.
This label-free training paradigm obviates the requirement for channel identification during inference and promotes generalization to previously unencountered channels, facilitated by the model’s reliance on acoustic features to guide adjustments and its exposure to a diverse set of channel conditions during training.

\begin{table*}[t]
\caption{CER (\%) and  Relative Improvement Rate (\%) of $\mathit{DEFA}$ on HAT.}
\label{tab:enc-dec-mismatch-hat}
\centering
\small\selectfont
\setlength{\tabcolsep}{3.8pt}
\begin{tabular}{l l
  *{8}{r r} r r
}
\toprule
\multirow{2}{*}{Method} 
  & \multirow{2}{*}{Channel}
  & \multicolumn{2}{c}{COND}   
  & \multicolumn{2}{c}{ADR}    
  & \multicolumn{2}{c}{ZM-X}
  & \multicolumn{2}{c}{ZM-Y}
  & \multicolumn{2}{c}{IPH}
  & \multicolumn{2}{c}{LAV}
  & \multicolumn{2}{c}{PCM}
  & \multicolumn{2}{c}{WCAM}
  & \multicolumn{2}{c}{AVG}
  \\
\cmidrule(lr){3-4}
\cmidrule(lr){5-6}
\cmidrule(lr){7-8}
\cmidrule(lr){9-10}
\cmidrule(lr){11-12}
\cmidrule(lr){13-14}
\cmidrule(lr){15-16}
\cmidrule(lr){17-18}
\cmidrule(lr){19-20}
& & CER & rel. & CER & rel. & CER & rel. & CER & rel. & CER & rel. & CER & rel. & CER & rel. & CER & rel. & CER & rel. \\
\midrule
$\mathit{{Van}_{pre}}$ & LAV & 1.21 & -- & 1.68 & -- & 1.68 & -- & 1.92 & -- & 1.64 & -- & 1.35 & -- & 1.34 & -- & 3.44 & -- & 1.78 & -- \\
$\mathit{{Van}_{adp}}$ & LAV & 1.22 & -0.8 & 1.49 & 11.3 & 1.48 & 11.9 & 1.64 & 14.6 & 1.44 & 12.2 & 1.27 & 5.9 & 1.27 & 5.2 & 2.95 & 14.2 & 1.60 & 10.1 \\
$\mathit{DEFA}$ & LAV & \textbf{0.97} & 19.8 & \textbf{1.22} & 27.4  & \textbf{1.18} & 29.8 & \textbf{1.27} & 33.9 & \textbf{1.14} & 30.5 &  \textbf{1.02} & 24.4 & \textbf{1.01} & 24.6 & \textbf{2.50} & 27.3 & \textbf{1.29} & 27.5 \\
\midrule
$\mathit{{Van}_{pre}}$ & ZM-X & 1.22 & -- & 1.53 & -- & 1.51 & -- & 1.65 & -- & 1.43 & -- & 1.27 & -- & 1.30 & -- & 2.90 & -- & 1.60 & -- \\
$\mathit{{Van}_{adp}}$ & ZM-X & 1.20 & 1.6 & 1.72 & -12.4 & 1.36 & 9.9 & 1.44 & 12.7 & 1.32 & 7.7 & 1.25 & 1.6 & 1.25 & 3.8 & 2.45 & 15.5 & 1.50 & 6.3 \\
$\mathit{DEFA}$ & ZM-X & \textbf{1.00} & 18.0 & \textbf{1.17} & 23.5 & \textbf{1.11} & 26.5 & \textbf{1.21} & 26.7 & \textbf{1.10} & 23.1 & \textbf{1.01} & 20.5 & \textbf{1.01} & 22.3 & \textbf{2.21} & 23.8 & \textbf{1.23} & 23.1 \\

\midrule
$\mathit{{Van}_{pre}}$ & \raisebox{-0.7ex}{\textasciitilde}WCAM & 1.04 & -- & 1.29 & -- & 1.27 & -- & 1.38 & -- & 1.25 & -- & 1.12 & -- & 1.08 & -- & 2.48 & -- & 1.36 & -- \\
$\mathit{{Van}_{adp}}$ & \raisebox{-0.7ex}{\textasciitilde}WCAM & 1.03 & 1.0 & 1.14 & 11.6 & 1.13 & 11.0 & 1.20 & 13.0 & 1.13 & 9.6 & 1.03 & 8.0 & 1.05 & 2.8 & 2.17 & 12.5 & 1.24 & 8.8 \\
$\mathit{DEFA}$ & \raisebox{-0.7ex}{\textasciitilde}WCAM & \textbf{0.90} & 13.5 & \textbf{0.97} & 24.8 & \textbf{0.97} & 23.6 & \textbf{1.01} & 26.8 & \textbf{0.93} & 25.6 & \textbf{0.92} & 17.9 & \textbf{0.93} & 13.9 & \textbf{1.70} & 31.5 & \textbf{1.04} & 23.5 \\

\bottomrule
\end{tabular}
\vspace{-10pt}
\end{table*}

\section{Experimental Setup}

\subsection{Datasets}

To conduct a comprehensive evaluation of the proposed methodology's efficacy, experiments were performed utilizing two benchmark datasets.

HAT \cite{liao2023}: The HAT corpus comprises approximately 1,461 hours of speech data, featuring utterances simultaneously recorded by eight distinct microphones.
This setup ensures identical speaker and linguistic content across channels, while varying the recording conditions.
Consequently, approximately 182.6 hours of audio data is available for each channel.
The recording devices encompass an \textbf{iPhone}, an \textbf{Android phone}, a \textbf{Webcam}, a professional \textbf{Condenser} microphone, a \textbf{Lavalier} microphone, a standard PC microphone (\textbf{PC-Mic}), and an X-Y stereo microphone (\textbf{ZOOM-X} and \textbf{ZOOM-Y}).

TAT \cite{liao2022}: To evaluate the robustness of the proposed channel encoder across different languages and recording devices, experiments were also conducted using the TAT corpus.
The TAT corpus exhibits similarities to the HAT corpus; however, it omits recordings from the \textbf{Webcam} and \textbf{PC-Mic} channels.

\subsection{$\mathit{{Enc}_{adp}}$ Training Setup}

The $\mathit{{Whisper}_{small}}$ model serves as the foundational ASR system.
Training of the adapter-enhanced encoder ($\mathit{{Enc}_{adp}}$) is conducted using the HAT corpus, which offers parallel recordings of identical utterances across eight synchronized channels.
The adapter architecture adheres to the methodology presented in \cite{huang2023}, wherein two lightweight adapter modules are integrated into each Transformer \cite{vaswani2017} block of the encoder.

The condenser channel is selected as the clean reference, a decision informed by its consistent demonstration of superior ASR performance across all evaluated models, as detailed in \Cref{fig:channel-analysis}.
This observation suggests that the condenser channel provides high-quality, acoustically clean input conducive to optimal ASR performance.
Conversely, the webcam channel, which consistently exhibits among the poorest performance metrics due to its substantial acoustic deviations from other channels, is designated as an unseen test condition. This allows for a rigorous evaluation of the model's generalization capabilities to challenging and acoustically distinct recording environments.
During the training phase, audio data from seven of the eight available channels are utilized, with the webcam channel explicitly excluded as the unseen condition.
The $\mathit{{Enc}_{adp}}$ module is trained for three epochs, employing a batch size of 24 and an initial learning rate of $10^{-4}$.
The AdamW optimizer \cite{loshchilov2018} is utilized, in conjunction with a linear learning rate scheduler that incorporates a warm-up phase corresponding to 10\% of the total training iterations.
Model checkpoints are selected based on optimal performance observed on the development set.

\begin{table*}[t]
\caption{CER (\%) and  Relative Improvement Rate (\%) of $\mathit{{Van}_{pre}}$ vs. $\mathit{{Van}_{adp}}$ on TAT.}
\label{tab:tat-result}

\centering
\small\selectfont
\setlength{\tabcolsep}{7pt}
\begin{tabular}{l l
  *{8}{r r} r r
}
\toprule
\multirow{2}{*}{Method}
  & \multirow{2}{*}{Channel}
  & \multicolumn{2}{c}{COND}
  & \multicolumn{2}{c}{ADR}
  & \multicolumn{2}{c}{ZM-X}
  & \multicolumn{2}{c}{ZM-Y}
  & \multicolumn{2}{c}{IPH}
  & \multicolumn{2}{c}{LAV}
  & \multicolumn{2}{c}{AVG}
  \\
\cmidrule(lr){3-4}
\cmidrule(lr){5-6}
\cmidrule(lr){7-8}
\cmidrule(lr){9-10}
\cmidrule(lr){11-12}
\cmidrule(lr){13-14}
\cmidrule(lr){15-16}
& & CER & rel. & CER & rel. & CER & rel. & CER & rel. & CER & rel. & CER & rel. & CER & rel. \\
\midrule
$\mathit{{Van}_{pre}}$ & COND & 8.92 & -- & 10.67 & -- & 10.97 & -- & 11.41 & -- & 9.80 & -- & 8.95 & -- & 10.12 & -- \\
$\mathit{{Van}_{adp}}$ & COND & \textbf{8.89} & 0.3 & \textbf{10.61} & 0.6 & \textbf{10.74} & 2.1 & \textbf{11.25} & 1.4 & \textbf{9.65} & 1.5 & \textbf{8.87} & 0.9 & \textbf{10.00} & 1.2 \\
\midrule
$\mathit{{Van}_{pre}}$ & ADR & 8.75 & -- & 10.28 & -- & 10.62 & -- & 11.15 & -- & 9.50 & -- & 8.87 & -- & 9.86 & -- \\
$\mathit{{Van}_{adp}}$ & ADR & \textbf{8.74} & 0.1 & \textbf{10.20} & 0.8 & \textbf{10.47} & 1.4 & \textbf{10.92} & 2.1 & \textbf{9.44} & 0.6 & \textbf{8.75} & 1.4 & \textbf{9.75} & 1.1 \\

\midrule
$\mathit{{Van}_{pre}}$ & ZM-X & 8.85 & -- & 10.38 & -- & 10.77 & -- & 11.13 & -- & 9.56 & -- & \textbf{8.85} & -- & 9.92 & -- \\
$\mathit{{Van}_{adp}}$ & ZM-X & \textbf{8.82} & 0.3 & \textbf{10.29} & 0.9 & \textbf{10.43} & 3.2 & \textbf{11.00} & 1.2 & \textbf{9.49} & 0.7 & 8.77 & -0.9 & \textbf{9.80} & 1.2 \\

\midrule
$\mathit{{Van}_{pre}}$ & ZM-Y & 9.04 & -- & 10.48 & -- & 10.68 & -- & 11.07 & -- & 9.63 & -- & \textbf{8.97} & -- & 9.98 & -- \\
$\mathit{{Van}_{adp}}$ & ZM-Y & \textbf{8.98} & 0.7 & \textbf{10.46} & 0.2 & \textbf{10.46} & 2.1 & \textbf{11.05} & 0.2 & \textbf{9.59} & 0.4 & 9.00 & -0.3 & \textbf{9.92} & 0.6 \\

\midrule
$\mathit{{Van}_{pre}}$ & IPH & 8.81 & -- & 10.41 & -- & 10.66 & -- & 11.24 & -- & 9.50 & -- & 8.83 & -- & 9.91 & -- \\
$\mathit{{Van}_{adp}}$ & IPH & \textbf{8.75} & 0.7 & \textbf{10.25} & 1.5 & \textbf{10.43} & 2.2 & \textbf{10.98} & 2.3 & \textbf{9.49} & 0.1 & \textbf{8.80} & 0.3 & \textbf{9.78} & 1.3 \\

\midrule
$\mathit{{Van}_{pre}}$ & LAV & 8.89 & -- & 10.77 & -- & 11.12 & -- & 11.66 & -- & 9.82 & -- & \textbf{8.89} & -- & 10.20 & -- \\
$\mathit{{Van}_{adp}}$ & LAV & \textbf{8.81} & 0.9 & \textbf{10.67} & 0.9 & \textbf{10.87} & 2.2 & \textbf{11.42} & 2.1 & \textbf{9.70} & 1.2 & \textbf{8.89} & 0.0 & \textbf{10.06} & 1.4 \\

\midrule
$\mathit{{Van}_{pre}}$ & \raisebox{-0.7ex}{\textasciitilde}ZM-Y & 8.47 & -- & 10.05 & -- & 10.24 & -- & 10.65 & -- & 9.15 & -- & 8.50 & -- & 9.51 & -- \\
$\mathit{{Van}_{adp}}$ & \raisebox{-0.7ex}{\textasciitilde}ZM-Y & \textbf{8.40} & 0.8 & \textbf{9.80} & 2.5 & \textbf{9.93} & 3.0 & \textbf{10.39} & 2.4 & \textbf{8.99} & 1.7 & \textbf{8.43} & 0.8 & \textbf{9.32} & 0.2 \\

\bottomrule
\end{tabular}
\vspace{-10pt}
\end{table*}

\subsection{Experiment Definitions}

This subsection delineates the notation and experimental configurations employed in our evaluations, facilitating a clear distinction between various encoder and decoder arrangements and the datasets utilized for fine-tuning.
Subsets of datasets incorporating specific channels are denoted by subscripts where appropriate.
For instance, subsets of the HAT and TAT datasets are represented as $\text{HAT}_{\text{COND}}$ (comprising only the condenser channel) and $\text{HAT}_{\raisebox{-0.8ex}{\textasciitilde}\text{WCAM}}$ (encompassing all channels except the webcam channel).
The prefix \raisebox{-0.7ex}{\textasciitilde} signifies ``exclusion'', and channel abbreviations conform to those presented in \Cref{fig:channel-analysis}.

To assess the efficacy of $\mathit{{Enc}_{adp}}$, we compare it to the pre-trained encoder ($\mathit{{Enc}_{pre}}$) by performing inference with each encoder combined with the same decoder.
We denote this vanilla inference setup as $Van_{enc} \mid \text{Data}$, where $enc \in \{Van_{pre}, Van_{adp}\}$ indicates the encoder used at inference time, and Data specifies the dataset which the decoder was fine-tuned on (always using outputs from $\mathit{{Enc}_{pre}}$).
For example, $Van_{adp} \mid \text{HAT}_{\text{COND}}$ refers to inference using the adapted encoder $\mathit{{Enc}_{adp}}$ with a decoder fine-tuned on the condenser channel subset of the HAT dataset.

To further explore the upper bound of $\mathit{{Enc}_{adp}}$, we introduce Decoder-Encoder Feature Adaptation ($\mathit{DEFA}$), which is specifically designed to adapt the decoder to the output distribution of $\mathit{{Enc}_{adp}}$.
In this procedure, $\mathit{{Enc}_{adp}}$ is combined with a decoder, and only the decoder is fine-tuned on the same dataset that was originally used for its fine-tuning.
This allows the decoder to adjust to the adapted encoder’s representations.
Since the vanilla encoder $\mathit{{Enc}_{pre}}$ already matches the decoder’s training distribution, no more adaptation is needed.
We denote this setup as $\mathit{DEFA} \mid \text{Data}$, following the same notation convention as the vanilla inference setup.
Both vanilla and $\mathit{DEFA}$ setups follow the same training configuration as $\mathit{{Enc}_{adp}}$, including training for three epochs, with model selection based on development set performance.
Since the vanilla decoders have already converged, this comparison is not significantly affected by differences in decoder fine-tuning duration, even if $\mathit{DEFA}$ decoders are fine-tuned for longer.

\section{Results}

\subsection{Main Results on HAT}

A potential challenge when applying $\mathit{{Enc}_{adp}}$ is that modifications to encoder outputs may introduce a mismatch with the decoder, potentially degrading ASR performance.
To investigate how the benefits of cleaner representations compete with the detrimental effects of encoder-decoder mismatch, we evaluate our proposed encoder $\mathit{{Enc}_{adp}}$ across a variety of decoders, each trained exclusively on data from a single channel.
As illustrated in the upper half of \Cref{tab:hat-result}, simply substituting $\mathit{{Enc}_{pre}}$ with $\mathit{{Enc}_{adp}}$ yields substantial improvements---this includes the decoder fine-tuned on the webcam channel, which remained unseen during the training of $\mathit{{Enc}_{adp}}$.
This indicates that our encoder generalizes well not only across channels but also across decoder configurations unseen during training.
These substantial improvements can be attributed to two factors: First, $\mathit{{Enc}_{adp}}$ preserves most of the linguistic and structural information in the original features, thereby limiting the degree of encoder-decoder mismatch.
Second, by effectively removing channel-related variations, $\mathit{{Enc}_{adp}}$ produces cleaner and more consistent feature representations, enabling decoders to achieve better performance.

Although minor degradations are observed in a few rare cases (e.g., $\mathit{{Van}_{adp}} \mid \text{HAT}_{\text{COND}}$ test on the condenser channel), overall performance consistently improves for the decoder, confirming the net benefit of applying $\mathit{{Enc}_{adp}}$.
Notably, even when $\mathit{{Enc}_{adp}}$ is applied to the condenser channel---the target domain of normalization---mismatch can still arise, since the normalized features remain approximations rather than exact replicas of real condenser data.
As demonstrated in subsequent experiments, fine-tuning the decoder to achieve better alignment with $\mathit{{Enc}_{adp}}$ mitigates this residual mismatch, resulting in more pronounced improvements.

Another notable advantage of our approach is its robust generalization capabilities under previously unseen conditions.
Even when evaluated on the unseen webcam channel, our method achieves relative improvements of approximately 10\% or greater across the majority of decoders, showcasing consistent effectiveness.
This remarkable performance in unseen scenarios underscores the generalization capability of $\mathit{{Enc}_{adp}}$.

To investigate the necessity of explicit normalization, we compare our approach against a strong baseline where the decoder is fine-tuned on data across multiple channels.
This configuration enables the decoder to directly observe channel variations during training, thereby raising the question of whether normalization still provides added value.
As demonstrated in the lower half of \Cref{tab:hat-result}, our method consistently yields significant improvements.
This confirms that encoder-side normalization provides complementary benefits, even with a decoder trained on diverse channels.

To isolate the effect of channel normalization, we avoid comparing $\mathit{{Enc}_{adp}}$ with an encoder fine-tuned on Hakka using ASR loss, as such a comparison would conflate normalization with language adaptation.
While our method is extensible via a Hakka-specific teacher encoder, we leave such language-aware adaptations to future work.

\begin{table*}[t]
\caption{CER (\%) and Relative Improvement Rate (\%) of $\mathit{DEFA}$ on TAT.}
\label{enc-dec-mismatch-tat}
\centering
\small\selectfont
\setlength{\tabcolsep}{7pt}
\begin{tabular}{l l
  *{6}{r r} r r
}
\toprule
\multirow{2}{*}{Method} 
  & \multirow{2}{*}{Channel}
  & \multicolumn{2}{c}{COND}   
  & \multicolumn{2}{c}{ADR}    
  & \multicolumn{2}{c}{ZM-X}
  & \multicolumn{2}{c}{ZM-Y}
  & \multicolumn{2}{c}{IPH}
  & \multicolumn{2}{c}{LAV}
  & \multicolumn{2}{c}{AVG}
  \\
\cmidrule(lr){3-4}
\cmidrule(lr){5-6}
\cmidrule(lr){7-8}
\cmidrule(lr){9-10}
\cmidrule(lr){11-12}
\cmidrule(lr){13-14}
\cmidrule(lr){15-16}
& & CER & rel. & CER & rel. & CER & rel. & CER & rel. & CER & rel. & CER & rel. & CER & rel. \\
\midrule
$\mathit{{Van}_{pre}}$ & ZM-X & 8.85 & -- & 10.38 & -- & 10.77 & -- & 11.13 & -- & 9.56 & -- & 8.85 & -- & 9.92 & -- \\
$\mathit{{Van}_{adp}}$ & ZM-X & 8.82 & 0.3 & 10.29 & 0.9 & 10.43 & 3.2 & 11.00 & 1.2 & 9.49 & 0.7 & 8.77 & 0.9 & 9.80 & 1.2 \\
$\mathit{DEFA}$ & ZM-X & \textbf{8.65} & 2.3 & \textbf{9.94} & 4.2 & \textbf{9.89} & 8.2 & \textbf{10.36} & 6.9 & \textbf{9.11} & 4.7 & \textbf{8.58} & 3.1 & \textbf{9.42} & 5.0 \\
\bottomrule
\vspace{-15pt}
\end{tabular}
\end{table*}

\begin{figure*}[ht]
    \centering
    \begin{subfigure}[b]{0.48\textwidth}
        \includegraphics[width=\linewidth]{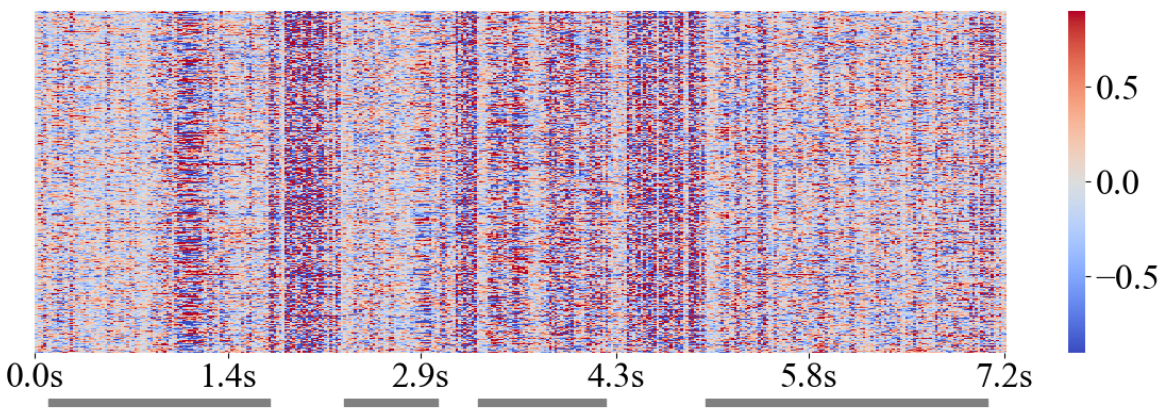}
    \end{subfigure}
    \hfill
    \begin{subfigure}[b]{0.48\textwidth}
        \includegraphics[width=\linewidth]{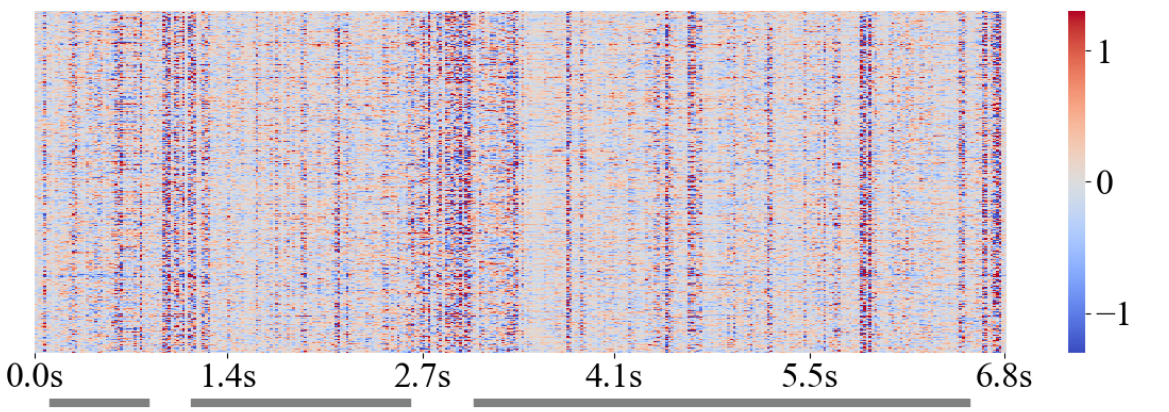}
    \end{subfigure}

    \vspace{2pt} 

    \begin{subfigure}[b]{0.48\textwidth}
        \includegraphics[width=\linewidth]{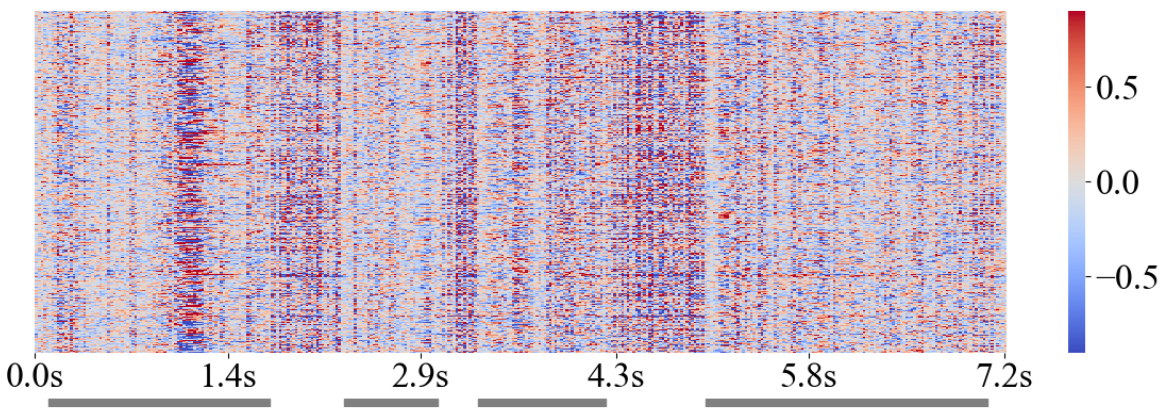}
        \caption{$\mathit{{Enc}_{pre}}$ (Top), $\mathit{{Enc}_{adp}}$ (Bottom), HAT dataset}
    \end{subfigure}
    \hfill
    \begin{subfigure}[b]{0.48\textwidth}
        \includegraphics[width=\linewidth]{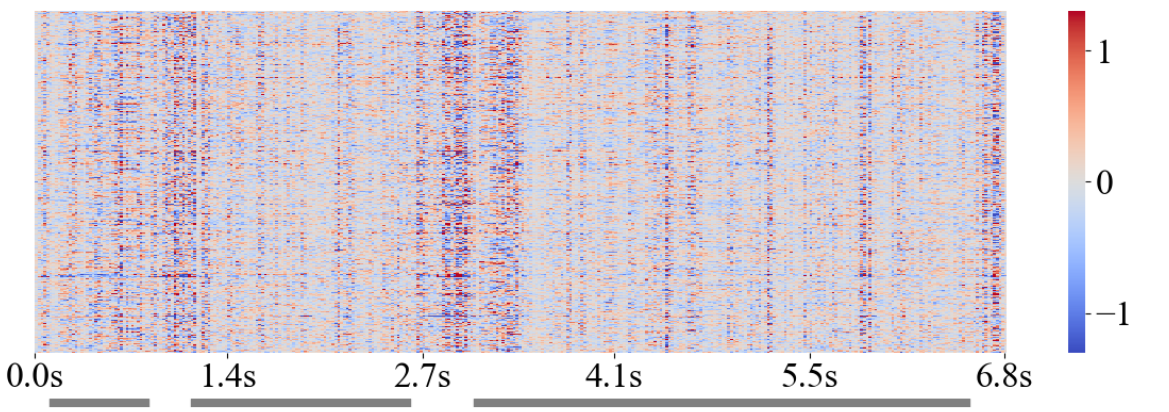}
        \caption{$\mathit{{Enc}_{pre}}$ (Top), $\mathit{{Enc}_{adp}}$ (Bottom), TAT dataset}
    \end{subfigure}
    \caption{
        Heatmap comparison of feature differences between condenser and Android phone channels.
        Lighter colors indicate smaller feature-level differences. 
        Top row: differences computed with $\mathit{{Enc}_{pre}}$. Bottom row: differences computed with $\mathit{{Enc}_{adp}}$. 
        Left column: HAT dataset. Right column: TAT dataset. 
        Gray lines below each heatmap indicate speech-active regions.
    }
    \label{fig:features-analysis}
    \vspace{-15pt}
\end{figure*}

\subsection{Encoder-Decoder Mismatch Analysis}

To better understand the full potential of channel normalization technology once encoder-decoder mismatch is addressed, we apply $\mathit{DEFA}$ to three decoders, each fine-tuned on a separate dataset: $\text{HAT}_{\text{LAV}}$, $\text{HAT}_{\text{ZM-X}}$, and $\text{HAT}_{\raisebox{-0.8ex}{\textasciitilde}\text{WCAM}}$.
The first two were selected to represent varying levels of encoder-decoder mismatch: $\text{HAT}_{\text{LAV}}$ shows minor mismatch, while $\text{HAT}_{\text{ZM-X}}$ reflects a more severe case--based on the average relative improvements shown in \Cref{tab:hat-result}.
Additionally, $\text{HAT}_{\raisebox{-0.8ex}{\textasciitilde}\text{WCAM}}$ is included as a strong baseline decoder trained on multi-channel data, representing a different scenario where the decoder is already exposed to channel variability. 
The results, shown in \Cref{tab:enc-dec-mismatch-hat}, demonstrate that across all conditions exhibit more significant improvements than $\mathit{{Van}_{adp}}$, with average relative gains exceeding 23\%.
Notably, cases where $\mathit{{Van}_{adp}}$ previously led to performance degradation are reversed into substantial gains---for example, $\mathit{DEFA} \mid \text{HAT}_{\text{ZM-X}}$ tested on the Android phone channel.
The results demonstrate that after further reducing the encoder-decoder mismatch, the channel normalization technique achieves even stronger improvements in ASR performance across different decoders.

\subsection{Language and Device Analysis on TAT}
To evaluate our method on decoders trained in other languages, we conducted experiments on TAT.
We tested decoders fine-tuned per channel and a strong multi-channel baseline, as in previous experiments, selecting the noisiest single-channel ZOOM-Y as an unseen case for the strong baseline.
Since the two datasets were collected at different times and likely with different devices, we treat TAT as a cross-device dataset.

As shown in \Cref{tab:tat-result}, $\mathit{{Enc}_{adp}}$ continues to help decoders even fine-tuned on different languages, with all tested models showing improvements, including the strong baseline.
With further reduction of encoder-decoder mismatch, our method continues to achieve more significant and comprehensive improvements across all test channels, including those that previously showed performance degradation (Lavalier), as shown in \Cref{enc-dec-mismatch-tat}.
Since $\mathit{DEFA}$ has already shown consistent gains across diverse decoder settings on HAT, here we evaluate its cross-lingual generalization by applying one channel.
Though the gains are less substantial than those on HAT, our method improves consistent performance across devices, demonstrating its cross-lingual and cross-device generalization capability.

\subsection{Features Visualization}
To investigate whether $\mathit{{Enc}_{adp}}$ normalizes features across different channels towards the condenser channel, we visualize the feature discrepancies between the condenser and another channel by utilizing a test sample.
We employ the Android phone channel as a representative example; however, analogous patterns are observed across other channels.

\Cref{fig:features-analysis} shows heatmaps of encoder output differences (condenser vs. Android phone), with lighter colors indicating greater similarity.
In the top row, the two heatmaps depict the difference in encoder outputs generated by $\mathit{{Enc}_{pre}}$ for the identical utterance recorded through both condenser and Android phone channels.
Conversely, the bottom row highlights the contrast between the outputs of $\mathit{{Enc}_{pre}}$ for the condenser input and $\mathit{{Enc}_{adp}}$ for the Android phone input.

We observe that our encoder reduces feature differences not only in speech regions but also in non-speech (background or silent) segments, indicating improved normalization across all acoustic contexts.
This effect is evident across both HAT (left) and TAT (right) datasets, where the feature gap between Android phone and condenser channels is effectively narrowed.
These findings align with the ASR results, demonstrating that our method enhances channel robustness at both the representation and task levels.


\section{Conclusion}
In this study\footnote{Code: \url{https://github.com/610494/channel-asr}.}, we clarify a common misconception by revealing that channel characteristics significantly contribute to ASR performance degradation, beyond the usual training-test mismatch explanation.
To address this, we propose a novel normalization technique that effectively mitigates channel-induced distortions and can be seamlessly integrated into existing pre-trained ASR models.
Our plug-and-play encoder adaptation enables easy replacement of the encoder to achieve strong channel robustness, with optional fine-tuning further boosting performance.
This approach improves ASR reliability across diverse recording conditions, facilitating more consistent and practical deployment in real-world applications.

\bibliographystyle{IEEEtran}
\bibliography{references}

\begin{thebibliography}{10}
\providecommand{\url}[1]{#1}
\csname url@samestyle\endcsname
\providecommand{\newblock}{\relax}
\providecommand{\bibinfo}[2]{#2}
\providecommand{\BIBentrySTDinterwordspacing}{\spaceskip=0pt\relax}
\providecommand{\BIBentryALTinterwordstretchfactor}{4}
\providecommand{\BIBentryALTinterwordspacing}{\spaceskip=\fontdimen2\font plus
\BIBentryALTinterwordstretchfactor\fontdimen3\font minus \fontdimen4\font\relax}
\providecommand{\BIBforeignlanguage}[2]{{%
\expandafter\ifx\csname l@#1\endcsname\relax
\typeout{** WARNING: IEEEtran.bst: No hyphenation pattern has been}%
\typeout{** loaded for the language `#1'. Using the pattern for}%
\typeout{** the default language instead.}%
\else
\language=\csname l@#1\endcsname
\fi
#2}}
\providecommand{\BIBdecl}{\relax}
\BIBdecl

\bibitem{gulati2020}
A.~Gulati, J.~Qin, C.-C. Chiu, N.~Parmar, Y.~Zhang, J.~Yu, W.~Han, S.~Wang, Z.~Zhang, Y.~Wu, and R.~Pang, ``Conformer: convolution-augmented transformer for speech recognition,'' in \emph{Proc. {Interspeech}}, 2020.

\bibitem{abouelenin2025}
A.~Abouelenin, A.~Ashfaq, A.~Atkinson, H.~Awadalla, N.~Bach, J.~Bao, A.~Benhaim, M.~Cai, V.~Chaudhary, C.~Chen, D.~Chen, D.~Chen, J.~Chen, W.~Chen, Y.-C. Chen, Y.-l. Chen, Q.~Dai, X.~Dai, R.~Fan, M.~Gao, M.~Gao, A.~Garg, A.~Goswami, J.~Hao, A.~Hendy, Y.~Hu, X.~Jin, M.~Khademi, D.~Kim, Y.~J. Kim, G.~Lee, J.~Li, Y.~Li, C.~Liang, X.~Lin, Z.~Lin, M.~Liu, Y.~Liu, G.~Lopez, C.~Luo, P.~Madan, V.~Mazalov, A.~Mitra, A.~Mousavi, A.~Nguyen, J.~Pan, D.~Perez-Becker, J.~Platin, T.~Portet, K.~Qiu, B.~Ren, L.~Ren, S.~Roy, N.~Shang, Y.~Shen, S.~Singhal, S.~Som, X.~Song, T.~Sych, P.~Vaddamanu, S.~Wang, Y.~Wang, Z.~Wang, H.~Wu, H.~Xu, W.~Xu, Y.~Yang, Z.~Yang, D.~Yu, I.~Zabir, J.~Zhang, L.~L. Zhang, Y.~Zhang, and X.~Zhou, ``Phi-4-{Mini} {Technical} {Report}: {Compact} yet powerful multimodal language models via mixture-of-{LoRAs},'' in \emph{Arxiv preprint {arXiv}:2503.01743}, 2025.

\bibitem{zhang2022}
Y.~Zhang, D.~S. Park, W.~Han, J.~Qin, A.~Gulati, J.~Shor, A.~Jansen, Y.~Xu, Y.~Huang, S.~Wang, Z.~Zhou, B.~Li, M.~Ma, W.~Chan, J.~Yu, Y.~Wang, L.~Cao, K.~C. Sim, B.~Ramabhadran, T.~N. Sainath, F.~Beaufays, Z.~Chen, Q.~V. Le, C.-C. Chiu, R.~Pang, and Y.~Wu, ``{BigSSL}: {Exploring} the frontier of large-scale semi-supervised learning for automatic speech recognition,'' \emph{IEEE Journal of Selected Topics in Signal Processing}, vol.~16, no.~6, pp. 1519--1532, 2022.

\bibitem{baevski2020}
A.~Baevski, Y.~Zhou, A.~Mohamed, and M.~Auli, ``wav2vec 2.0: {A} framework for self-supervised learning of speech representations,'' in \emph{Proc. {NeurIPS}}, 2020.

\bibitem{radford2023}
A.~Radford, J.~W. Kim, T.~Xu, G.~Brockman, C.~Mcleavey, and I.~Sutskever, ``Robust speech recognition via large-scale weak supervision,'' in \emph{Proc. {ICML}}, 2023.

\bibitem{chan2021}
W.~Chan, D.~Park, C.~Lee, Y.~Zhang, Q.~Le, and M.~Norouzi, ``{SpeechStew}: {Simply} mix all available speech recognition data to train one large neural network,'' in \emph{Arxiv preprint {arXiv}: 2104.02133}, 2021.

\bibitem{zhang2023}
Y.~Zhang, W.~Han, J.~Qin, Y.~Wang, A.~Bapna, Z.~Chen, N.~Chen, B.~Li, V.~Axelrod, G.~Wang, Z.~Meng, K.~Hu, A.~Rosenberg, R.~Prabhavalkar, D.~S. Park, P.~Haghani, J.~Riesa, G.~Perng, H.~Soltau, T.~Strohman, B.~Ramabhadran, T.~Sainath, P.~Moreno, C.-C. Chiu, J.~Schalkwyk, F.~Beaufays, and Y.~Wu, ``Google {USM}: {Scaling} automatic speech recognition beyond 100 languages,'' in \emph{Arxiv preprint {arXiv}: 2303.01037}, 2023.

\bibitem{shraddha2022}
S.~Shraddha, J.~L. G, and S.~K. S, ``Child speech recognition on end-to-end neural {ASR} models,'' in \emph{Proc. {CONIT}}, 2022.

\bibitem{wang_dysarthric_2024}
H.~Wang, Z.~Jin, M.~Geng, S.~Hu, G.~Li, T.~Wang, H.~Xu, and X.~Liu, ``Enhancing pre-trained {ASR} system fine-tuning for dysarthric speech recognition using adversarial data augmentation,'' in \emph{Proc. {ICASSP}}, 2024.

\bibitem{qin2021}
Y.~Qin, W.~Liu, Z.~Peng, S.-I. Ng, J.~Li, H.~Hu, and T.~Lee, ``Exploiting pre-trained {ASR} models for {Alzheimer}'s disease recognition through spontaneous speech,'' in \emph{Arxiv preprint {arXiv}: 2110.01493}, 2021.

\bibitem{zusag2024}
M.~Zusag, L.~Wagner, and B.~Thallinger, ``{CrisperWhisper}: {Accurate} timestamps on verbatim speech transcriptions,'' in \emph{Proc. {Interspeech}}, 2024.

\bibitem{wang2025}
C.-C. Wang, L.-W. Chen, C.-K. Chou, H.-S. Lee, B.~Chen, and H.-M. Wang, ``Channel-aware domain-adaptive generative adversarial network for robust speech recognition,'' in \emph{Proc. {ICASSP}}, 2025.

\bibitem{mani2020}
A.~Mani, S.~Palaskar, N.~V. Meripo, S.~Konam, and F.~Metze, ``{ASR} error correction and domain adaptation wsing machine translation,'' in \emph{Proc. {ICASSP}}, 2020.

\bibitem{zhou2021}
K.~Zhou, Y.~Yang, Y.~Qiao, and T.~Xiang, ``Domain generalization with {MixStyle},'' in \emph{Proc. {ICLR}}, 2021.

\bibitem{liao2019}
C.-F. Liao, Y.~Tsao, H.-Y. Lee, and H.-M. Wang, ``Noise adaptive speech enhancement using domain adversarial training,'' in \emph{Proc. {Interspeech}}, 2019.

\bibitem{schroter2022}
H.~Schröter, A.~Maier, A.~Escalante-B, and T.~Rosenkranz, ``Deepfilternet2: {Towards} real-time speech enhancement on embedded devices for full-band audio,'' in \emph{Proc. {IWAENC}}, 2022.

\bibitem{defossez2020}
A.~Défossez, G.~Synnaeve, and Y.~Adi, ``Real time speech enhancement in the waveform domain,'' in \emph{Proc. {Interspeech}}, 2020.

\bibitem{pandey2019}
A.~Pandey and D.~Wang, ``{TCNN}: {Temporal} convolutional neural network for real-time speech enhancement in the time domain,'' in \emph{Proc. {ICASSP}}, 2019.

\bibitem{ochiai2024}
T.~Ochiai, K.~Iwamoto, M.~Delcroix, R.~Ikeshita, H.~Sato, S.~Araki, and S.~Katagiri, ``Rethinking processing distortions: disentangling the impact of speech enhancement errors on speech recognition performance,'' \emph{IEEE/ACM Trans. Audio, Speech, Lang. Process.}, 2024.

\bibitem{iwamoto2022}
K.~Iwamoto, T.~Ochiai, M.~Delcroix, R.~Ikeshita, H.~Sato, S.~Araki, and S.~Katagiri, ``How bad are artifacts?: {Analyzing} the impact of speech enhancement errors on {ASR},'' in \emph{Proc. {Interspeech}}, 2022.

\bibitem{wang2024}
K.-C. Wang, Y.-J. Li, W.-L. Chen, Y.-W. Chen, Y.-C. Wang, P.-C. Yeh, C.~Zhang, and Y.~Tsao, ``Bridging the {Gap}: {Integrating} pre-trained speech enhancement and recognition models for robust speech recognition,'' in \emph{Proc. {EUSIPCO}}, 2024.

\bibitem{ho2023}
K.-H. Ho, E.-L. Yu, J.-W. Hung, and B.~Chen, ``{NAaLOSS}: {Rethinking} the objective of speech enhancement,'' in \emph{Proc. {MLSP}}, 2023.

\bibitem{iwamoto2024}
K.~Iwamoto, T.~Ochiai, M.~Delcroix, R.~Ikeshita, H.~Sato, S.~Araki, and S.~Katagiri, ``How does end-to-end speech recognition training impact speech enhancement artifacts?'' in \emph{Proc. {ICASSP}}, 2024.

\bibitem{houlsby2019}
N.~Houlsby, A.~Giurgiu, S.~Jastrzebski, B.~Morrone, Q.~D. Laroussilhe, A.~Gesmundo, M.~Attariyan, and S.~Gelly, ``Parameter-efficient transfer learning for {NLP},'' in \emph{Proc. {ICML}}, 2019.

\bibitem{hu2023}
Z.~Hu, L.~Wang, Y.~Lan, W.~Xu, E.-P. Lim, L.~Bing, X.~Xu, S.~Poria, and R.~Lee, ``{LLM}-{Adapters}: {An} adapter family for parameter-efficient fine-tuning of large language models,'' in \emph{Proc. {EMNLP}}, 2023.

\bibitem{sung2022}
Y.-L. Sung, J.~Cho, and M.~Bansal, ``{VL}-{Adapter}: {Parameter}-efficient transfer learning for vision-and-language tasks,'' in \emph{Proc. {CVPR}}, 2022.

\bibitem{liao2023}
Y.-F. Liao, S.-H. Hwang, Y.-S. Chen, H.-C. Lai, Y.-H. Chung, L.-T. Shen, Y.-C. Huang, C.-J. Huang, H.~W. Han, L.-W. Chen, P.-C. Su, and C.-S. Huang, ``Taiwanese {Hakka} across {Taiwan} corpus and {Formosa} speech recognition challenge 2023 - {Hakka} {ASR},'' in \emph{Proc. {O}-{COCOSDA}}, 2023.

\bibitem{liao2022}
Y.-F. Liao, J.~S. Tsay, P.~Kang, H.-L. Khoo, L.-K. Tan, L.-C. Chang, U.-G. Iunn, H.-L. Su, T.-G. Thiann, H.-K. Tiun, and S.-L. Liao, ``Taiwanese across {Taiwan} corpus and its applications,'' in \emph{Proc. {O}-{COCOSDA}}, 2022.

\bibitem{huang2023}
Z.~Huang, H.~Xing, and M.~Liu, ``Adapter {Integration}: {Mitigating} catastrophic forgetting in multi-language and multi-accent whisper {ASR} model fine-tuning,'' 2023.

\bibitem{vaswani2017}
A.~Vaswani, N.~Shazeer, N.~Parmar, J.~Uszkoreit, L.~Jones, A.~N. Gomez, L.~Kaiser, and I.~Polosukhin, ``Attention is all you need,'' in \emph{Proc. {NeurIPS}}, 2017.

\bibitem{loshchilov2018}
I.~Loshchilov and F.~Hutter, ``Decoupled weight decay regularization,'' in \emph{Proc. {ICML}}, 2018.

\end{thebibliography}
\end{document}